\documentclass[a4paper,aps,pra,twocolumn,showpacs,
]{revtex4}

\usepackage{amsfonts,amssymb,amsmath}         
\usepackage{graphics,graphicx,epsfig}         
\usepackage{bm}
\usepackage{color}

\newcommand{\beq}{\begin{equation}}
\newcommand{\eeq}{\end{equation}}
\newcommand{\beqa}{\begin{eqnarray}}
\newcommand{\eeqa}{\end{eqnarray}}
\newcommand{\ket} [1] {\vert #1 \rangle}
\newcommand{\bra} [1] {\langle #1 \vert}

\newcommand{\mean}[1]{\langle #1 \rangle}

\begin{document}

\title{On parametric type interaction between light and atomic ensembles}
\author{V.N. Gorbachev, A.I. Zhiliba, A.A. Rodichkina, A.I.
Trubilko}

\affiliation
{Laboratory of Quantum Information $\&$ Computation,\\
University of AeroSpace Instrumentation, 67, Bolshaya Morskaya,
St.-Petersburg, 190000, Russia}


\begin{abstract}
One-photon and Raman type interactions between two-level atoms and narrow-band light
are considered. We give some exactly solvable models of these processes when only one-photon
Fock states are involved in the evolution. Possible application of these models for generation
and transformation of entangled states of the W-class, some of which demonstrate hierarchy
structure, are discussed. Finally, we consider preparation of entangled chains of atomic ensembles.

\end{abstract}


\maketitle


\section{Introduction}

Interaction between light and atoms still represents a promising
way of preparation of multiparticle entangled states to be main
resource of many quantum communications, in particularly, a
recently introduced one-way quantum computer \cite{OWQC},
\cite{OWQC1}. In this Letter we focus on some of the informational
tasks which can be realized by considering a model in which an
ensemble of two-level atoms interacts as a single
quantum-mechanical system with one photon light. The multiparticle
problem of a such type, also known as Dicke one \cite{Dicke}, in
general, has no exact solutions and has been studied under various
approximations, particularly, in the master equation for atoms
with bath-like radiation field \cite{DickeMasterEq} and also using
the second-order many-body perturbation theory
\cite{DickeS-Opapprox}.

One of the main obstacle in obtaining of exact solutions in this
model is a large number of degrees of freedom of the multiparticle
system. However, we could expect that only a little part of them
will play a role, if we consider interaction with a small number
of photons. Indeed, if we use, say, one-photon Fock states as
initial, hence there is a small number of excitations in the
system, evolution of which now can be obtained directly by making
use the time-evolution operator method. It is derived in the
Appendix the solutions of a general type for evolution of the wave
functions of the system N atoms and quantum field when a few
photons only are involved in the parametric type interaction.

These solutions describe coherent processes of exchange of excitations between
light and the atomic ensemble and look like three-photon parametric phenomena
well studied in Nonlinear Optics. In this way the symmetric states, belonging
to the Dicke family with a subset of well known in the quantum informational
theory W entangled states \cite{CiracW}, can be generated. Using these solutions
we focus on preparation and transformation of the W states that can accomplish
many informational tasks such as telecloning \cite{Clon},
\cite{Tclon}, quantum cryptography \cite{LeeSecret}, multiparticle teleporation,
dense coding \cite{Wchan} and probabilistic teleportation \cite{TelWJap}, \cite{TelWCorea}.
For these states the distillation protocol and its optical implementation have been
introduced \cite{DistilWcn}. The analitical solutions for single photons interacted
with three level atoms of the lambda configuration have been found based on arguments\cite{Fleishh}
associated with electromagnetically induced transparency (EIT) and considered as a
basis for storage of light. In the case of the squeezed field storage or quantum
memory for light has been proposed and demonstrated experimentally \cite{Ptheor},
\cite{Pexp}.
The paper is organized as follows. First we consider two
hamiltonians that describe one-photon and Raman interactions
between light and atoms. Then we introduce the necessary notations
and features of the symmetric and W-states and write some exact
solutions for the problems. Finally, we discuss generation and
conversion of the states from the W-class some of which have a
hierarchic organization, such as considered in ref. \cite{VAAW},
and it is assumed to be actual in quantum bioinformatics \cite{MAl}.

\section{Hamiltonians}
Now we consider multiphoton interactions between $N$ two-level
identical atoms and narrow-band light which the frequency range
and detuning from resonance are small with respect to a decay rate
of the atomic levels, $\gamma$. Assuming the time evolution is not
long in the sense that $t\ll \gamma^{-1}$, atomic relaxation can
be neglected and hence the resonance approximation becomes valid.
An effective Hamiltonian can be used to describe a set of various
phenomena, particularly multiphoton processes among of which we
focus only on two. The first is a one-photon resonance interaction
of a single mode $a$. The second is the Raman type interaction of
two modes $b,c$ whose frequencies obey the relation
$\omega_{b}-\omega_{c}=\omega_{0}$, where $\omega_{0}$ is a
frequency of an atomic transition. Both processes can be modelled
by the following Hamiltonians
\begin{eqnarray}
\label{001} 
H_{1}=i\hbar g(aS_{10}-a^{\dagger}S_{01}),&&\\
\label{002} H_{II}=i\hbar
f(c^{\dagger}bS_{10}-cb^{\dagger}S_{01}),&&
\end{eqnarray}
where the atomic operators are defined by the notation
$S_{xy}=\sum_{a}s_{xy}(a), s_{xy}(k)=\ket{x}_{a}\bra{y}, x,y=0,1$,
$\ket{0}_{a}$ and $\ket{1}_{a}$ are lower and upper levels of a
single atom, $Z=a,b,c$ are annihilation operators of a photon at
frequency $\omega_{Z}$ and $g,f$ are real coupling constants.
\\
Because the Hamiltonians commute with the operator of permutation of particles,
symmetry of the input wave function is unchanged during evolution, and any
symmetric states of the Dicke family, particularly W-states, can
be generated. There is a simple physical reason for this. When $m$ atoms
(from $N$) absorb photons, then in the case of identical particles, all
possibilities to do this result in a symmetric superposition state that consists
of any states with $m$ excited atoms.

\section{Symmetric Dicke and W states}

The family of the symmetric Dicke states \cite{Dicke} can describe an
ensemble of $N$ identical two-level atoms in which any $m<N$
particles are excited. In general, they read
\begin{eqnarray}
\label{003}
\ket{m;N}=\sum_{z}P_{z}\ket{\underbrace{1,\dots,1}_{m},
\underbrace{0,\dots,0}_{N-m}},&&
\end{eqnarray}
where $P_{z}$ is one from $C_{m}^{N}=N!/(m!(N-m)!)$
distinguishable permutations of particles. In (\ref{003}) the vectors
are not normalized and $ \mean{m;N|m;N}=C_{m}^{N}.$ Using the dipole momentum
operator $S_{10}=\sum_{a}^{N}s_{10}(a)$ one finds that
\begin{equation}\label{004}
S_{10}^{m}\ket{0}=m!\ket{m;N}.
\end{equation}
This representation tells us, that the symmetric states can be produced
in any process involving interactions of collective dipole
momentum of the atomic ensemble. One of the possible examples is given
by the Hamiltonians (\ref{001}), (\ref{002}). If $m=1$, then
\begin{eqnarray}
 \label{005} \nonumber \ket{1;N}
=\ket{1,0,0,\dots,0}+\ket{0,1,0,\dots,0}&&\\
+\ket{0,0,1,\dots,0}+\ket{0,0,0,\dots,1}.&&
\end{eqnarray}
In quantum information theory the normalized vector
$1/\sqrt{N}\ket{1;N}=W$ is well known as a multiparticle $W$ state.
The symmetrical states (\ref{003}) are entangled over many
criteria and any pairs of particles are entangled in contrast with
Greenberger-Horne-Zeilinger states in which two particles are separable.

The following properties of these states will be further used delow
\begin{eqnarray}
\label{006} \nonumber
S_{01}\ket{0;N}=0,&&\\
\nonumber
 S_{10}\ket{m;N}=(m+1)\ket{m+1;N},&&\\
 \nonumber
 S_{01}\ket{m;N}=(N-m+1)\ket{m-1;N},&&\\
 \nonumber
 S_{01}S_{10}\ket{m;N}=(m+1)(N-m)\ket{m;N},&&\\
 S_{10}S_{01}\ket{m;N}=m(N-m+1)\ket{m;N}.&&
\end{eqnarray}

\section{Some exact solutions}

The multiparticle problems of interaction between light and atoms
given by the Hamiltonians (1) and (2 )have a set of exact solutions
for particular cases of initial field states. For example, consider that
atoms are illuminated by light in the one-photon Fock states $\ket{n},
\ket{01},\ket{10}$, $n=0,1$. In this way, one finds a simple
obvious evolution for the Hamiltonian (\ref{001})
\begin{eqnarray}
\label{3001} \nonumber \exp\{-i\hbar^{-1}H_{1}t\}
\Big\{\alpha\ket{1}\otimes\ket{0;N}+\beta\ket{0}\otimes\ket{1;N}\Big\}&&\\
\nonumber =
\alpha\Big\{c~\ket{1}\otimes\ket{0;N}+s~\frac{1}{\sqrt{N}}
\ket{0}\otimes\ket{1;N}\Big\}&&\\
+\beta\Big\{-s~\sqrt{N}\ket{1}\otimes
\ket{0;N}+c~\ket{0}\otimes\ket{1;N}\Big\},&&
\end{eqnarray}
where $c,s=\cos\theta, \sin\theta and \theta=tg\sqrt{N}$. It worth noting that
Eq. (\ref{3001}) describes exchange of one excitation between light
and atoms similar to a beamsplitter with a single photon as its input
\begin{eqnarray}
\label{3002} \nonumber
 \alpha\ket{10}+\beta\ket{01} \to
\alpha\Big\{c\ket{10}+s\ket{01}\Big\}&&\\
+\beta\Big\{-s\ket{10}+c\ket{01}\Big\}.&&
\end{eqnarray}
It is well known, that a beamsplitter can be modelled by an
effective Hamiltonian of the form $H_{BS}=i\hbar
g(a^{\dagger}b-h.c.)$, where two modes of field $a$ and $b$ have
equal frequencies but differ in their wave vectors or ``which
path``. Indeed, the beamsplitter Hamiltonian also describes a
process of frequency conversion in a strong classical pump
$\omega_{a}-\omega_{b}=\Omega$, where two modes $a$ and $b$ have
different frequencies or (''color'') and $\Omega$ is a frequency of the
pump. The process belongs to a family of the three-photon
parametric phenomena well known in Quantum Optics. We then see
that Eq. (\ref{3001}) looks as the beamsplitter-like solution and
can describe a parametric process. From the point of view of three-photon
parametric phenomena the Hamiltonian (\ref{002}) could be treated as parametric
interaction involving two photons of the light modes $c$ and $b$ and the third
''photon'' represents an atomic ensemble:
\begin{eqnarray}
\label{0022} H_{II}=i\hbar
f(c^{\dagger}bS-cb^{\dagger}S^{\dagger}),
\end{eqnarray}
where we introduce the notation $S=S_{10}, S^{\dagger}=S_{01}$. In
this case one finds
\begin{eqnarray}
\label{3004}
 \nonumber
\exp\{-i\hbar^{-1}H_{II}t\}(\alpha\ket{01}+\beta\ket{10})\otimes\ket{\psi}_{A}&&\\
\nonumber = \alpha\Big\{\cos\Big[tf\sqrt{S^{\dagger}S}\Big]\ket{01}&&\\
\nonumber
 +S\frac{1} {\sqrt{S^{\dagger}S}}\sin\Big[
tf\sqrt{S^{\dagger}S}\Big]\ket{10}\Big\}
\nonumber
\otimes\ket{\psi}_{A} &&\\
+\beta\Big\{-S^{\dagger}\frac{1}{\sqrt{SS^{\dagger}}}\sin\Big[tf\sqrt{
SS^{\dagger}}\Big]\ket{01}&&\\
\nonumber
+\cos\Big[tf\sqrt{SS^{\dagger}}\Big]\ket{10}\Big\}
 \otimes\ket{\psi}_{A},&&
\end{eqnarray}
where $\ket{\psi}_{A}$ is an atomic wave function. If we choose the
atomic state as $\ket{\psi}_{A}=\ket{m;N}$, then using (\ref{006}) we have
\begin{eqnarray}
\label{3005}
 \nonumber
(\alpha\ket{01}
+\beta\ket{10})\otimes\ket{m;N}\to
\nonumber \alpha\Big\{\cos\theta_{m}\ket{01}\otimes\ket{m;N}&&\\
\nonumber
 +\sqrt{\frac{m+1}{N-m}}
\sin\theta_{m}\ket{10}\otimes\ket{m+1;N}\Big\}&&\\
\nonumber
+\beta\Big\{-\sqrt{\frac{N-m+1}{m}}\sin\theta'_{m}\ket{01}\otimes
\ket{m-1;N}&&\\
 +\cos\theta'_{m}\ket{10}\otimes\ket{m;N}\Big\}&&
\end{eqnarray}
where $\theta_{m}=tf\sqrt{(m+1)(N-m)}$,
$\theta'_{m}=tf\sqrt{m(N-m+1)}$. It follows from the presented solution (\ref{3005}),
that during the exchange of a single excitation between two modes and atoms,
a set of atomic symmetric states is produced.
Both the wave vectors given by (\ref{3001}) and (\ref{3005})
describe entangled states of light and atoms. However this
multipartite entanglement has a hierarchic organization consisting
itself from the atomic W-states. Note that the similar W-hierarchy of
the atomic states can be prepared also by a projection measurement,
introduced in \cite{VAAW}.

\section{Generation and transformation of W states}

One of the main features of the state conversions given by the
solutions (\ref{3001}) and (\ref{3005}) is swapping. It results in
a set of processes for the preparation of multiparticle entanglement,
their transformation and storage. Possible schemes could be
implemented in two ways. First is a cavity version containing an atomic
ensemble which interacts with high-Q cavity modes and the required initial
state is injected. In the second scheme, a trapped atomic ensemble is
illuminated by light from a single photon source. It follows from (\ref{3005})
that there are two processes of transformation at least. The first is generation
of a W state: $\ket{0;N}\to\ket{1;N}$, where one atom is excited only. There
is also an opposite process of decreasing of the excited atoms: $
\ket{1;N}\to\ket{0;N}$. It can be treated as a disentanglement
operation. Generally, if an ensemble with $m$ excited atoms are
prepared, then transformation of the form $\ket{m;N}\to\ket{m\pm1;N}$
can be achieved.

Both solutions, given by (\ref{3001}) and (\ref{3005}), describe an
swapping between light and atoms. This is a basis for quantum
memory, when a state of light, particularly an unknown state, can
be stored in an atomic ensemble. For example, examining (\ref{3001}) we find
\begin{eqnarray}
\nonumber \Big(\alpha\ket{1}+\beta\ket{0}\Big)\otimes\ket{0;N}&&\\
\to \ket{0}\otimes
\Big\{\alpha\frac{1}{\sqrt{N}}\ket{1;N}+\beta\ket{0;N}\Big\}.
\end{eqnarray}
As well it follows from (\ref{3005}), that storage can be realized for
entangled states of light $\alpha\ket{01}+\beta\ket{10}$. The same possibility
has been remarked in case of an ensemble of atoms in a coupled lambda type
configuration and assuming adiabatic approximation \cite{Fleishh}.

The beamsplitter-like solution (\ref{3001}) results in a scheme
that prepares entanglement of the W class from atomic ensembles.
The scheme is an analog of the version with two
beamsplitters \cite{VAAW}. Let us consider a consecutive interaction
between light in a single photon state and two atomic ensembles
having $N_{1}$ and $N_{2}$ atoms. Then the output state of
light and the first ensemble can be input for the next interaction.
Using (\ref{3001}) we find
\begin{eqnarray}
\nonumber
 \ket{1}\otimes \ket{0;N_{1}}\otimes\ket{0;N_{2}}\to&&\\
 \nonumber
 \Big(c_{1}\ket{1}\otimes\ket{0;N_{1}}+s_{1}\ket{0}\otimes
 (1/\sqrt{N_{1}})\ket{1;N_{1}}\Big)\otimes\ket{0;N_{2}}&&\\
\nonumber
\to c_{1}c_{2}\ket{1}\otimes\ket{0;N_{1}}\otimes\ket{0;N_{2}}&&\\
\nonumber
+c_{1}s_{2}(1/\sqrt{N_{2}}\ket{0}\otimes\ket{0;N_{1}}\otimes\ket{1;N_{2}}&&\\
+s_{1}(1/\sqrt{N_{1}}\ket{0}\otimes\ket{1;N_{1}}\otimes\ket{0;N_{2}}.&&
\end{eqnarray}
As a result, a three-partite entanglement of the W class is prepared.
Indeed, the desired consecutive interaction can be done, for
example, by the Stark effect. Let two trapped atomic samples in free
space be illuminated by light and two static fields are applied
to the working levels. Then, by manipulating the static fields, it
is possible to manipulate the resonance interaction between light
and any of the atomic ensembles.

Another version of the entanglement preparation is possible when
we consider interaction between light and a set of atomic
ensembles. Let ensembles be located at the points $x=a_{1},a_{2},\dots
a_{\nu}$. We will call this configuration an atomic chain. Let the
chain be illuminated by light so that its interaction is described
by the Hamiltonian (\ref{0022}), where operators $S$ are now the sum
of operators of atomic ensembles, that are located in
$x=a_{1},\dots,a_{\nu}$ and have $N_{1},\dots,N_{\nu}$ atoms
\begin{eqnarray}
S=\sum_{x}S_{10}(x)
\end{eqnarray}
For simplicity, we will neglect all spatial behavior of light
within the chain. Let $\ket{O}$ be initial state of atoms, its
ground state, then using (\ref{3004}), we have
\begin{eqnarray}
\label{4000}
 \nonumber
\ket{01}\otimes\ket{O}\to&&\\
\cos\theta_{0}\ket{01}\otimes\ket{O}+(1/\sqrt{N'})
\sin\theta_{0}\ket{10}\otimes \ket{\mathcal{C}}, 
\end{eqnarray}
where $\theta_{0}=tf\sqrt{N}$, $N'=N_{1}+\dots +N_{\nu}$, and the
vector $\ket{\mathcal{C}}$ is defined as
\begin{eqnarray}
\label{Chain} \nonumber
\ket{\mathcal{C}}=\sum_{x}S_{10}(x)\ket{O}
\nonumber
=\ket{1;N_{1}}\ket{0,\dots,0}&&\\
+\ket{0}\ket{1;N_{2}}\ket{0,\dots,0}+\dots+
\ket{0,\dots,0}\ket{1;N_{\nu}}.&&
\end{eqnarray}
From (\ref{Chain}) one finds the atomic chain to be in the $W$
state consisting of $\nu$ atomic ensembles.

\section{Conclusions}

In summary, we have discussed simple analytical solutions
describing the evolution of the wave functions for a system of N
atoms and quantum field when a few photons only are involved in
the parametric type interactions. From a theoretical point of
view, these solutions are interesting not only for their
remarkable simplicity, but also because they are one of the
simplest analytic solutions that describe a set of conversion
processes of the symmetric atomic states from W-class and can be
useful for modelling of generation and transformation of entangled
states, in particular, storing light states on atoms and as well
preparing atomic entangled chains both to be processes of
fundamental importance in quantum information.
We acknowledge discussions with A. Kazakov. One of us (AZ) would like to thank
M. Shapiro for useful remarks.

This work was supported in part by the Delzell Foundation Inc. and INTAS grant
no. 00-479.
\appendix
\section{}
The presented set of exact solutions aries from the next
observation on the Hamiltonian written in the form
\begin{eqnarray}\label{512}
\nonumber
H=i\hbar\vartheta&&\\
\upsilon =\pi^{\dagger} h-\pi h^{\dagger},&&
\end{eqnarray}
where there is one requirement only for operators, that
$[\pi;h]=[\pi;h^{\dagger}]=0$. Let two wave functions $\Phi$ and
$\Phi_{\dagger}$ obey the conditions:
\begin{eqnarray} \label{513}
\nonumber
h\Phi=A\Phi_{\dagger},\qquad h\Phi_{\dagger}=0&&\\
h^{\dagger}\Phi_{\dagger}=B\Phi, \qquad h^{\dagger}\Phi=0,&&
\end{eqnarray} where $A$ and $B$ are c-numbers or commuting operators, that
 commutate with $\pi,\pi^{\dagger}$: $[X;Y]=0, Y=A,B, X=\pi, \pi^{\dagger}$.
Then considering the initial state of the form
\begin{equation}\label{514}
    \varphi=c\Phi+e\Phi_{\dagger},
\end{equation}
one finds its evolution
\begin{eqnarray} \label{516}
\nonumber
 \exp\Big[\vartheta t\Big](c\Phi+e\Phi_{\dagger})
 \nonumber
 =c\Big\{\cos[t\sqrt{\pi\pi^{\dagger}AB}]\Phi&&\\
\nonumber
 +
\pi^{\dagger}\frac{1}{\sqrt{\pi\pi^{\dagger}A
B}}A\sin[t\sqrt{\pi\pi^{\dagger}AB}]\Phi_{\dagger}\Big\}&&\\
\nonumber + e\Big\{ -\pi\frac{1} {\sqrt{\pi^{\dagger}\pi B
 A}}B\sin[t\sqrt{\pi^{\dagger}\pi BA}]\Phi&&\\
 +
\cos[t\sqrt{\pi^{\dagger}\pi BA}]\Phi_{\dagger}\Big\}. &&
\end{eqnarray}
The Hamiltonian (\ref{512}) describes a family of physical processes
between light and atoms, including multiphoton parametric phenomena.
Then for appropriate input states one can find exact solution of the
form (\ref{516}). When $\pi^{\dagger}=g,\pi=g^{*}$, $h=S_{10}a^{M}$
we find $M$-photon absorbtion
\begin{eqnarray}
\label{50001} \vartheta=gS_{10}a^{M}-g^{*}S_{01}a^{\dagger M},&&
\end{eqnarray}
for which the exact solution of the form (\ref{516}) arises, where
$\Phi=\ket{2M-p}\otimes \ket{0;N},
\Phi_{\dagger}=\ket{M-p}\otimes\ket{1;N}$, $p=1,\dots,M$,
$A=A(p)=((2M-p)(2M-p-1)\dots(M-p))^{1/2}$, $B=A(p-1)$. A
particular case of $M=1$ reduces to the considered Hamiltonians
(\ref{001}), (\ref{002}) if $\pi^{\dagger}=S_{10}$, $h=ga,
fc^{\dagger}b$.
The presented arguments are true, in particularly, for three-photon parametric
processes well known in Nonlinear Optics
\begin{equation}\label{518}
\vartheta=f(a^{\dagger}bc-ab^{\dagger}c^{\dagger}).
 \end{equation}
The closed solution of the problem in general case are unknown for the authors,
however for particular case of input states $\Phi=\ket{0,1,n}$ and
$\Phi_{\dagger}=\ket{1,0, n-1}$, for which $A=B=\sqrt{n}$, we
obtain the exact solution
\begin{eqnarray} \label{519}
\nonumber
c\ket{0,1,n}+e\ket{1,0,n-1}\to
c\Big\{\ket{0,1,n}\cos[tf\sqrt{n}]&&\\
\nonumber
 +\ket{1,0,n-1}\sin[tf\sqrt{n}]\Big\}
&&\\
\nonumber
 +
e\Big\{-\ket{0,1,n}\sin[tf\sqrt{n}]+&&\\
\ket{1,0,n-1}\cos[tf\sqrt{n}]\Big\}.&&
\end{eqnarray}

\end{document}